\documentclass[twocolumn]{aastex701}
\usepackage{siunitx}
\newcommand{\tmag}{T$_{\rm mag}$}
\newcommand{\nlcs}{324,530}
\newcommand{\nstars}{185,422}

\begin{document}

\title{ExoVetter and a TESS Exoplanet Transit Search of Kepler Stars as Ground Truth}

\author[0000-0001-9692-9751]{Matthew M. Dallas}
\affiliation{Space Telescope Science Institute, 3700 San Martin Drive, Baltimore, MD 21218}
\email[show]{mdallas@stsci.edu}
\correspondingauthor{Matthew M. Dallas}

\author[0000-0001-7106-4683]{Susan E. Mullally}
\affiliation{Space Telescope Science Institute, 3700 San Martin Drive, Baltimore, MD 21218}
\email{smullally@stsci.edu}

\author[0009-0004-7656-2402]{Fergal Mullally}
\affiliation{Constellation, 1310 Point Street, Baltimore, MD 21231}
\email{fergal.mullally@gmail.com}

\author[0000-0001-9269-8060]{Michelle Kunimoto}
\affiliation{ Department of Physics and Astronomy, University of British Columbia, 6224 Agricultural Road, Vancouver, BC V6T 1Z1, Canada}
\email{mkuni@phas.ubc.ca}
\author[0000-0002-7127-7643]{Te Han} 
\affil{Department of Physics and Kavli Institute for Astrophysics and Space Research, Massachusetts Institute of Technology, Cambridge, MA 02139, USA}
\email{tehan@mit.edu} 

\author[0000-0003-2630-8073]{Timothy D.~Brandt}
\affiliation{Space Telescope Science Institute, 3700 San Martin Drive, Baltimore, MD 21218}
\email{tbrandt@stsci.edu}

\begin{abstract}

We introduce a Python package called \texttt{exoVetter} that produces vetting metrics and visualizations to evaluate transiting exoplanet signals found in high-cadence time-series photometry. \texttt{exoVetter} is open source, modular, and independent of the mission data so that it can be easily expanded and used as part of different transit search pipelines tuned to different science cases and missions. We demonstrate its use in a pipeline as a way to distinguish real events found in a box-least squares exoplanet transit search of \nlcs\ light curves from the Transiting Exoplanet Survey Satellite mission; we limit our demonstration to stars also observed by the more sensitive Kepler telescope as a way to establish ground truth for the events found in the light curves. Using this ground truth, we provide baseline thresholds for several flux-level vetters and one centroid-offset vetter when using \texttt{exoVetter} metrics on TESS data. Our demonstration also indicates that it is possible to create a planet catalog from the light curves of dim targets ($>$13.5\tmag) that has a similar fraction of real events to that of the bright targets ($<$13.5\tmag) for short period events identified in the TESS-Gaia light curves \citep{Han2023AJ}.

\end{abstract}


\section{Introduction} \label{sec:intro}

The transit method remains one of the most popular techniques for discovering and characterizing exoplanets, offering precise measurements of orbital periods and planetary radii \citep[e.g.,][]{Thompson2018, Guerrero2021, Christiansen2025}. Space-based surveys such as \textit{CoRoT} \citep{CoRoT}, \textit{Kepler} \citep{Borucki2010}, and the \textit{Transiting Exoplanet Survey Satellite} (\textit{TESS}; \citealt{Ricker2015}) have transformed this field by enabling uninterrupted, high-cadence, high-precision photometry of millions of stars. To date, \textit{TESS} has discovered over 900 planets that are now confirmed according to the NASA Exoplanet Archive \citep{Confirmed_planets}, with thousands of additional candidates awaiting confirmation. Upcoming missions such as \textit{Roman} and \textit{PLATO} \citep{PLATO2025, Wilson2023} will extend transit-based searches by obtaining time-series photometry of an estimated $\sim 59\times 10^{6}$ bright (F146$<$21 mag) stars across \textit{Roman}'s Galactic Bulge Time Domain Survey, and $\gtrsim 260,000$ bright (V$\leq$13 mag) stars/ $\gtrsim 5,000$ M dwarfs (V$\leq$16 mag) across \textit{PLATO}'s P1/P5 and P4 stellar samples, respectively, further increasing the scale of exoplanet detection efforts.

This exponential growth in light curve data has driven the development of automated pipelines for identifying and validating transit signals. A typical transit search pipeline involves three primary steps: (1) light curve extraction, (2) signal detection using algorithms such as the box-least squares periodogram \citep[BLS;][]{Kovacs2002}, and (3) vetting to classify signals as either planet candidates or false positives. While early exoplanet searches often relied on manual inspection, which was labor-intensive and subject to unmeasurable human biases, the scale of modern surveys necessitates automated and reproducible vetting approaches. 

The Kepler mission pioneered several automated and semi-automated vetting frameworks, including the \texttt{Robovetter} \citep{Coughlin2016, Thompson2018}, a rule-based decision tree classifier designed to emulate manual vetting heuristics, and the centroid robovetter \citep{Mullally2017ksci.rept...23M}, designed to look for evidence of background eclipsing binary blends. Kepler's final data release (DR25) used the \texttt{Robovetter} to produce a fully uniform planet catalog, characterized in terms of completeness and reliability, to support robust population studies. Complementary machine learning-based tools for Kepler such as the \texttt{Autovetter} \citep{McCauliff2015} and \texttt{SIDRA} \citep{sidra2016MNRAS.455..626M} were developed to automate candidate classification using training sets derived from earlier human-labeled catalogs. Since the beginning of TESS and K2 operations, several vetting pipelines and packages have been developed by the community to classify transiting exoplanets in the growing TESS data sets, such as \texttt{AstroNet-K2} \citep{Dattilo2019}, \texttt{DAVE} \citep{Kostov2019}, \texttt{EDI-Vetter} \citep{Zink2020}, \texttt{ExoMiner++} \citep{ExoMiner2025AJ....170..287V}, \texttt{LATTE} \citep{Eisner2020}, \texttt{LEO-Vetter} \citep{LEO-Vetter}, \texttt{RAVEN} \citep{RAVEN}, and \texttt{TRICERATOPS} \citep{TRICERATOPS}. While each package calculates its own selected metrics and plots to evaluate a signal, each also has its own interfaces and dependencies, making it difficult to use them together and compare results. Also, each additional package creates its own burden on the community to maintain, update, and expand as it is applied to future research projects.

In this work we present a Python-based software package for evaluating the possible transit signals produced by a typical planet search. \texttt{exoVetter}, described in \S\ref{sec:exovetter}, is open source, modular, extensible and mission agnostic. It is not an end-to-end classification pipeline, but a package of vetting metrics developed on a common set of base-level code with one unified interface that can be used to evaluate transit-like events. New metrics can be added to meet the vetting needs of new science cases or missions. In \S\ref{sec:search} we demonstrate the current workflow and performance of \texttt{exoVetter} by running it against a set of TESS light curves with known transit-like events as observed by the Kepler mission. The methods and software used to perform this search made use of compute at Amazon Web Services (AWS) proximate to the TESS light curves hosted by MAST at AWS. In \S\ref{subsec:cloud} we describe the cloud-based workflow we used to run a transit search and classifier that makes use of \texttt{exoVetter}.  

\section{ExoVetter}
\label{sec:exovetter}
To simplify the calculation and evaluation of exoplanet vetting metrics we have developed a software package called \texttt{exoVetter}\footnote{\url{https://zenodo.org/records/19700753}}. This package provides a common Python interface to algorithms developed to vet transits, including those used by the \texttt{KeplerRobovetter}, \texttt{DAVE}, and \texttt{LEO-Vetter} \citep{Thompson2018, Kostov2019, LEO-Vetter}. The code is modular so future vetting algorithms can be included as the community develops them. By having them in one package they rely on a unified framework that makes use of shared utility and unit tests, which is simpler to maintain long term. Additionally, each vetting algorithm can be called independently to create a unique pipeline for a given science case. \texttt{exoVetter} also provides interactive plots to further explore the validity of the results for a specific ephemeris and light curve combination. \texttt{exoVetter} is open source and available on GitHub\footnote{\url{https://github.com/spacetelescope/exovetter}} or on PyPI\footnote{\url{https://pypi.org/project/exovetter/}} for the most recent stable release. 

\texttt{exoVetter} is built using popular software packages such as \texttt{lightkurve} \citep{lightkurve} on the front end to manipulate time series flux and time series images, and \texttt{numpy}, \texttt{astropy}, and \texttt{scipy} \citep{Harris2020, astropy2022, scipy} on the back end. Most vetting metrics only require a \texttt{lightkurve} object with time and flux, and an \texttt{exoVetter} Threshold Crossing Event (TCE) object containing the transit time, duration, depth and period. \texttt{exoVetter} includes helper functions commonly used for time series analysis of high cadence data such as phase folding, marking in-transit cadences, and converting between missions with different time offsets. \texttt{exoVetter} also includes a centroid offset vetter which requires the input of a target pixel time series object and a TCE object. The vetting tools are implemented directly in {\tt exoVetter} to avoid unnecessary dependencies on other packages, with only the locality preserving projections (LPP) vetter relying on the external \texttt{lpproj}\footnote{\url{https://github.com/jakevdp/lpproj}} package to perform LPP transformations.

Each vetter can be run separately, with or without plotting, allowing users to execute the pieces of the vetting that are most valuable for their use cases. For example, those doing automated vetting on millions of light curves may want to execute a bespoke set of vetting metrics without producing millions of plots. Then for those signals that pass the initial vetting, the same algorithm can be re-run with plotting turned on to inspect the results. Since plotting functions all make use of \texttt{matplotlib} \citep{Hunter2007}, the resulting plots can be made to be interactive with certain backends, or sent to a static output. \texttt{exoVetter} also includes several Jupyter notebooks demonstrating typical workflows.

The algorithms incorporated into \texttt{exoVetter} at this time are described in detail in the original papers describing the planet searches \citep{LEO-Vetter, Kostov2019}. We welcome additional well-tested and documented vetters to the package that will help it adapt to the needs of the science community searching for transit signals in long-baseline, high-cadence data sets. The \texttt{exoVetter} package includes several utility functions for dealing with time domain data and transit ephemerides that are used for many of the vetters. These include functions for calculating transit phases, the in-transit times, and constants used by well-known missions to offset their Julian dates.

\texttt{exoVetter} includes both vetters designed to distinguish transits from noise and data systematics and vetters to distinguish planetary transits from astrophysical false positives like background eclipsing binaries. In this section we summarize the flux and pixel vetters intended to address noise and systematics; these vetters are further demonstrated in \S\ref{sec:search} by applying them to TESS data. All current vetters, including those intended for astrophysical false positives, are described and demonstrated in the GitHub notebook tutorials. \texttt{exoVetter} does include vetters to separate eclipsing binaries from transit signals, namely odd-even and SWEET (\citealt{Thompson2018} Sections A.3.2, A.4.14). 

\begin{itemize}
\item \textbf{SHP (Transit Shape Test):}
Measures the relative strength of the deepest Model-Shift responses in the negative and positive directions; see Section 3.10 of \citet{LEO-Vetter}. Lower values (which are still $> 0$) are more transit-like, while higher values are increasingly dominated by brightening events.

\item \textbf{MES (Multiple Event Statistic):}
Uses the folded transit train described by the TCE to determine the significance of the event; see Section 2.1 of \citet{LEO-Vetter}. Higher values indicate the transit is less dominated by noise.

\item \textbf{DMM (Depth Mean-to-Median Ratio Test):}
Measures the ratio of the mean and median depths across the transit train; see Section 3.6 of \citet{LEO-Vetter}. Values closer to 1 indicate transit depths consistent with real planet transits.

\item \textbf{CHI (SNR Consistency Test):}
Compares the signal-to-noise ratio (SNR) of each transit against the overall weighted average transit depth; see Section 3.7 of \citet{LEO-Vetter}. Higher values indicate consistent transit depths across the transit train. 

\item \textbf{LPP (Locality Preserving Projections):}
Compares the shape of the folded light curve to the shape of known transits using the dimensionality reduction algorithm LPP and k-nearest neighbors; see \citet{lpp}. Values $>$~3 are typically not transit-like.

\item \textbf{ModShift (Model-Shift Uniqueness Test):}
Measures the significance of transit-like events in the folded light curve and compares them to the primary; see Section A.3.4 of \citet{Thompson2018}. For our purposes we use the $MS_2$ metric which compares the primary and tertiary events relative to the amount of noise in the light curve to determine if the phased event is unique. Higher values indicate an increasing probability the TCE is due to quasi-periodic noise.
\end{itemize}

The \texttt{exoVetter} methods also create useful plots for individual inspection of the TCE as measured by that light curve or image time series. We demonstrate the plots and how to interpret them in the next sections.

\subsection{VizTransit}
\begin{figure*}
\centering
\includegraphics[width=0.9\textwidth]{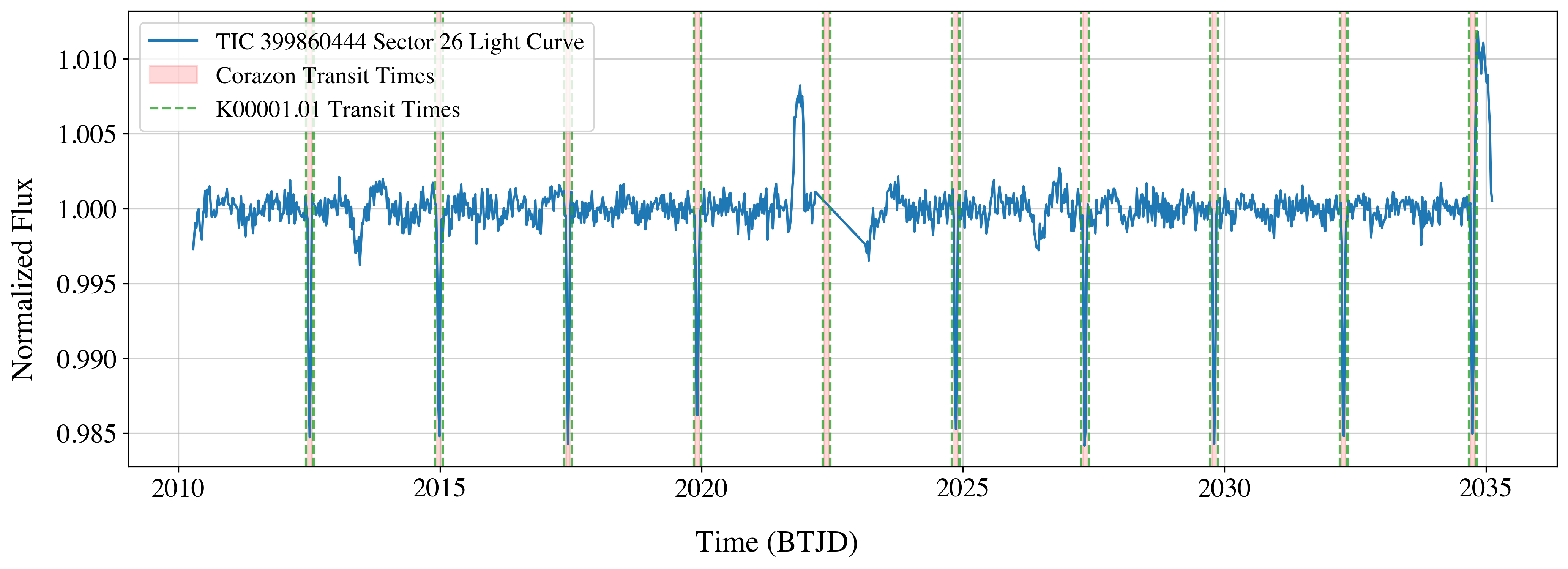}
\caption{TGLC calibrated aperture flux for TIC 399860444 (K00001.01) in TESS Sector 26. The confirmed transit times are overlaid in green while the federated TCE transit times discussed in \S\ref{sec:experiment} are overlaid in red. This \texttt{lightkurve} object and TCE are used as input to \texttt{VizTransit} to produce the individual and folded transits shown in Figure \ref{fig:viz_transit_example}. 
\label{fig:viz_transit_train}}
\end{figure*}

The \texttt{VizTransit} class can be used to quickly plot each individual transit along with the corresponding folded light curve. As with most of the vetters, its inputs are a TCE and a \texttt{lightkurve} object. We demonstrate a simple use case from the sample we discuss in \S\ref{sec:experiment} by running \texttt{VizTransit} on a TCE generated on the confirmed planet TrES-2 b (K00001.01, \citet{tres2}). For reference we first show the light curve searched with the compared TCE overlaid in Figure \ref{fig:viz_transit_train}.

\begin{figure*}
\centering
\includegraphics[width=0.9\textwidth]{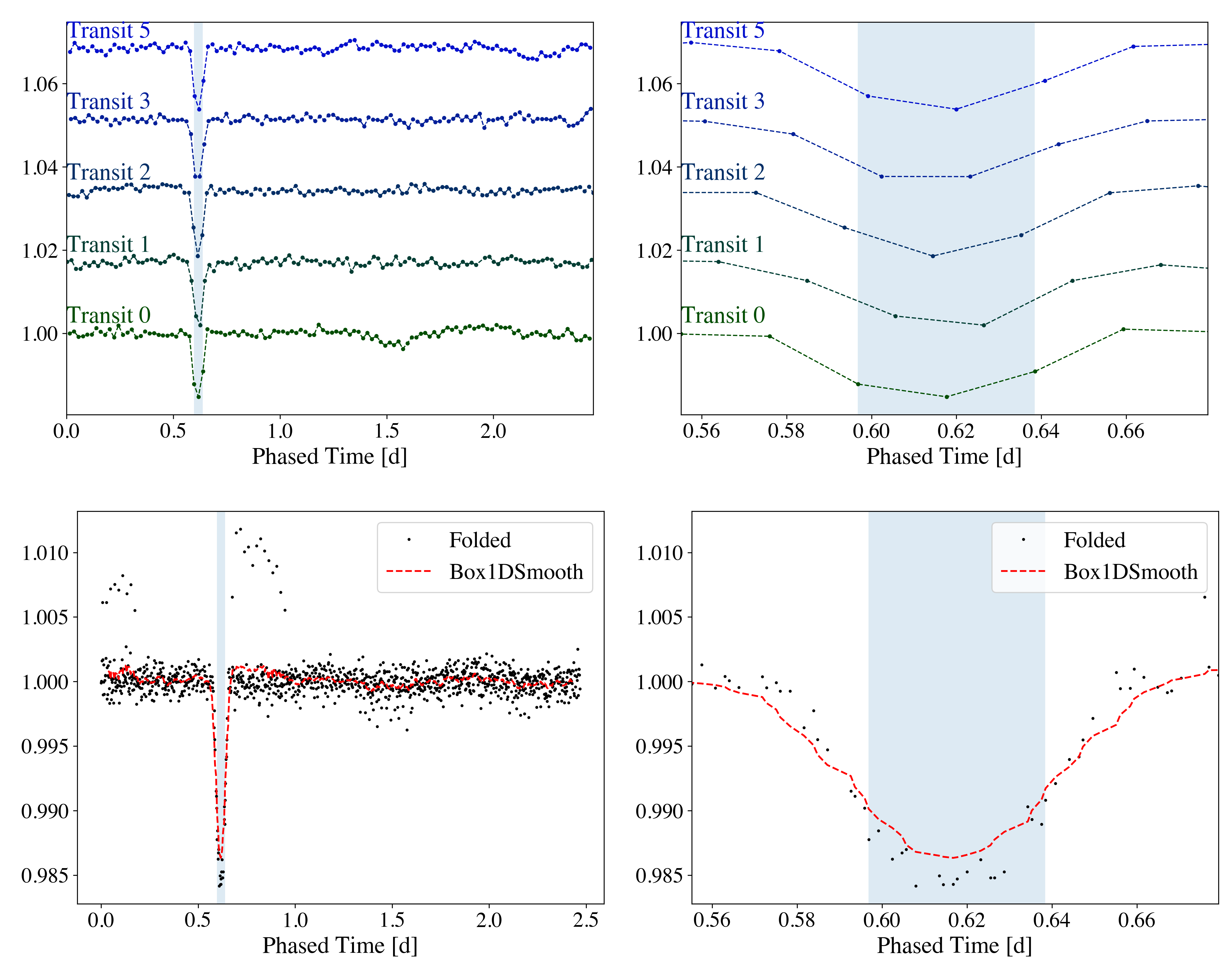}
\caption{The \texttt{VizTransit} outputs for K00001.01 with the right column set with \texttt{transit\_only = True} to zoom in on the transit, \texttt{max\_transits = 5} to limit to the first five individual transits on the top row, and \texttt{smooth = 3} to use three data points as the width in the smoothing filter in the bottom row. Note that \texttt{VizTransit} will automatically skip a transit that has no data (i.e., the fourth ``transit'' during the gap in Figure \ref{fig:viz_transit_train} is not shown and \texttt{VizTransit} correctly labels the next valid transit as ``5''). 
\label{fig:viz_transit_example}}
\end{figure*}

The first five transits along with all the transits folded as displayed by \texttt{VizTransit} are shown in Figure \ref{fig:viz_transit_example}. The right column is the same output but with the argument \texttt{transit\_only} set to \texttt{True} which centers the plot to three transit durations around the midpoint. The vetter also includes options to tune the smoothing filter or remove it in the folded plot, and to limit or increase the number of individual transits shown in the upper panels.

\subsection{Centroid Vetter}
The \texttt{Centroid} module in \texttt{exoVetter} is an implementation of the centroid vetter used by DAVE for the K2 mission. A more thorough description is provided in Section 2.1 of \citet{Kostov2019}. The goal of the vetter is to help determine the sky location of the observed transit signal and distinguish whether it is on the given target or is a result of blending with a nearby eclipsing binary signal. Using a TCE and \texttt{lightkurve} target pixel object, the centroid vetter calculates the difference image between the in-transit and out-of-transit images, then fits the centroids of the in-transit, out-of-transit and difference images. By default the vetter outputs the resulting offset measured from the difference centroid to the out-of-transit centroid location along with the statistical significance of the observed location. Following \citet{Kostov2019}, the centroid vetter averages the row and column offsets and assumes they follow 2D Gaussian distributions. The statistical significance is then calculated as the probability of observing the offset using \texttt{scipy}'s Rayleigh survival function. If the \texttt{plot} option is set to \texttt{True}, then probability density ellipses encapsulating the offsets at $68\%$, $95\%$, and $99.7\%$ confidence intervals are also drawn after the individual difference images. Difference images that have failed Gaussian fits or that are significantly different from the others in a stack are excluded from this calculation and are plotted in red in the significance plot. Significance values close to $1$ indicate the transit is consistent with the location of the target star and values less than $\sim 0.003$ suggest the target star is not the source of the transit. For an example of applying the centroid vetter and significance value to a real planet search, see \citet{Diego2026}.

\begin{figure*}
\centering
\includegraphics[width=0.7\textwidth]{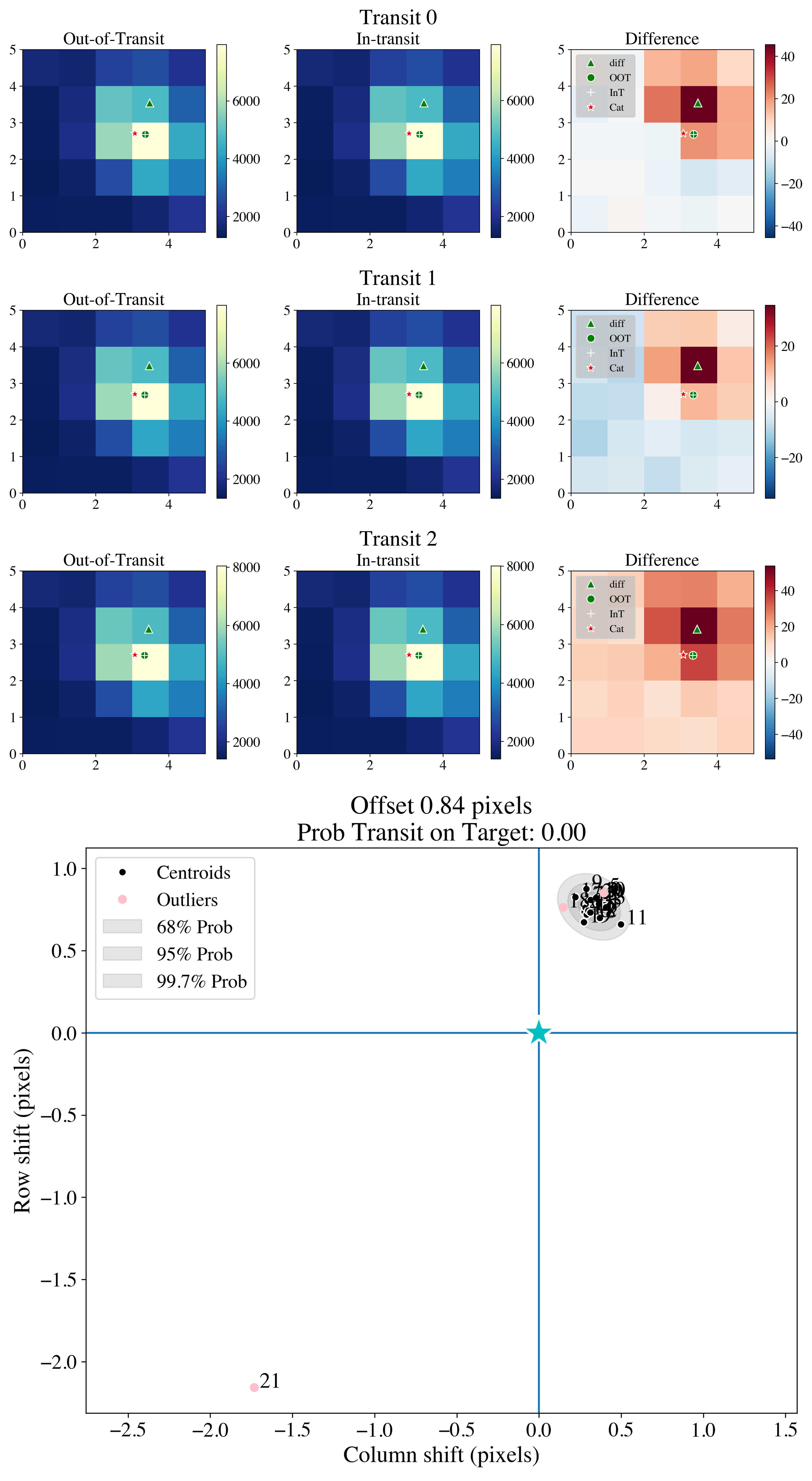}
\caption{The \texttt{Centroid} difference image and significance plot outputs for KOI 6642.01 with the centroid offset taken from the Gaia catalog position of the target available on MAST. For visual clarity we only show the first three transit difference images, although 22 transits occur over the sector. Each of the transits is represented in the significance plot shown beneath, with outliers plotted in red.
\label{fig:centroid_offset_example}}
\end{figure*}

The \texttt{exoVetter} implementation also includes options to measure the offset relative to a catalog position. This is especially helpful if there are other stars in the target pixel file brighter than the target star which may move the measurement of the out-of-transit position. In Figure \ref{fig:centroid_offset_example} we provide an example of the \texttt{Centroid}-produced plots which show a real centroid offset for TIC 184561543 in TESS Sector 15 that was identified by the Kepler team around the corresponding KOI K06642.01. We further discuss how the \texttt{Centroid} vetter performs on our demonstration search data in \S\ref{subsec:centroid}. 

\section{ExoVetter Demonstration} \label{sec:experiment} \label{sec:search}
We demonstrate \texttt{exoVetter} as part of a transit search using TGLC light curves. We chose to run our search and vetting pipeline on the TESS targets previously observed by Kepler to ensure we had ground truth to evaluate the performance of the vetters on known transit-like events.

\subsection{Kepler as Ground Truth} \label{subsec:KOIs}

Any pipeline that is used to detect and evaluate a transit signal needs to be evaluated to determine which transiting planet signals can be found using that pipeline on those data (detection efficiency) and which signals the pipeline detects by mistake (false alarms). While the detection efficiency is usually evaluated by injecting transit signals into known light curves \citep[e.g.,][]{2020AJ....160..159C}, measuring the frequency of false alarms through simulated data is more difficult since it requires knowledge of the ground truth about the frequency of all noise sources in the data. Fortunately, TESS observed hundreds of thousands of stars that were also observed by the Kepler space telescope. Since the noise levels of the Kepler light curves are much lower than for TESS, Kepler can act as a limited ground truth for evaluating the performance of our TESS transit detection pipeline. 

This method has its limitations. The Kepler survey did not search for transits around many stars brighter than 12th magnitude \citep{Thompson2018} and intentionally selected stars to be main-sequence FGK-type stars. Due to the size of the cameras, TESS's ability to find transits decreases substantially beyond 14th magnitude because of noise in the light curves \citep{Ricker2015}. The Kepler field is found in the portion of the TESS survey where most stars are only observed for one sector ($\approx$ 27 days) per year with few stars being observed in consecutive sectors. As a result, this search using the Kepler stars as ground truth is best for evaluating the performance of our pipeline on dim stars ($>$12 magnitude) and short period planets ($<$12 days). The total number of stars observed by Kepler we use to test our pipeline is \nstars\ (see Table \ref{tab:tglc_noise_transit}). Our ground truth is not perfect, and it is worth noting that our sample of true candidates is derived from the KOI table which includes eclipsing binaries, thereby contaminating our true positives. Additionally, the Kepler mission may have missed true transiting systems when generating their KOI catalog, though we believe this latter effect is small for short period transiting planets given the increased baseline and sensitivity of Kepler compared to TESS.

\subsection{Light Curves} \label{subsec:TESS}
We run our transit search of Kepler targets using TESS-Gaia light curves (TGLC).
\citet{Han2023AJ} create these light curves by leveraging Gaia DR3 astrometry to produce calibrated light curves for stars as faint as $T \sim 16$. Stars searched by Kepler were typically 12 -- 16 \tmag, so the extra depth provided by TGLC allows us to search many more stars than some of the other light curve products currently available. TGLC includes both PSF and aperture light curves. We use the PSF flux when available, switching to the calibrated aperture flux for the remaining targets. 

We identify \nstars\ targets in TESS sectors 14, 15, 16, and 26 (those which overlapped the Kepler field) with TGLC data products that match a star found in the Kepler Stellar catalog \citep{Mathur2017}. We create this list by cross-matching the Gaia IDs used by TGLC with Kepler IDs that are associated with a TIC ID in MAST \citep{Stassun2019}. For simplicity, our \nstars\ targets have any TICs assigned to multiple Gaia DR3 sources removed. This occurred when a single DR2 source (the basis of the TIC ID assignment) was resolved into multiple DR3 sources (the basis of TGLC) and only affected $<1\%$ of the original cross-matched sample. Across the TESS sectors we searched there were \nlcs\ TESS light curves created by the TESS Gaia Light Curve project on these targets common to both TESS and Kepler.

We then create a control sample out of the target list based on whether they are also listed by the Kepler Project in the Kepler Object of Interest (KOI) table \citep{KOI_DOI}. This control sample corresponds to targets with no associated KOIs and comprises a majority of our original dataset with 177,850 targets. The remaining sample includes 7,572 targets, each with any number of KOIs. A full breakdown of the targets and data products per sector is given in Table \ref{tab:tglc_noise_transit}. 

\subsection{Search and Vetting Pipeline} 
\label{subsec:BLS}

For this demonstration we adapt a simple planet transit search pipeline called Corazon, available on GitHub\footnote{\url{https://github.com/spacetelescope/corazon}}. Corazon implements a standard box-least squares transit search \citep{Kovacs2002} using Astropy's implementation of the algorithm \citep{astropy2022,astropy2018}. Prior to the search, Corazon removes cadences that are identified as poor quality in the TGLC pipeline, or those that are identified as generally bad across that sector according to the TESS data release notes\footnote{\url{https://archive.stsci.edu/tess/tess_drn.html}} due to scattered light or thermal events. We do not use the other detrending algorithms available in Corazon since the TGLC products are already detrended and normalized. The box-least squares algorithm performs the search using transit durations of 1, 2, 4, 8 and 12 hours for periods greater than 0.8 days. The search is iterative in that it removes the cadences associated with the in-transit points of any signal it finds and then searches again. The search terminates when either the fraction of data points remaining is less than 70\% of the original, or once the signal returned has an SNR less than 1. In Corazon, SNR is calculated as the model depth divided by the standard deviation of the median detrended light curve after the model was subtracted. For this search, we also set Corazon to return no more than 6 events from a single light curve. For each signal found by Corazon, we use the \texttt{exoVetter} package to return the suite of metrics discussed in 
\S\ref{sec:exovetter}.

We execute this search and vetting pipeline in the AWS cloud; \S\ref{subsec:cloud} provides details on the data and cloud engineering. From the no-KOI control sample and transit-like KOI reference sample targets we search 311,381 and 13,149 TGLC light curves, respectively. The number of TGLC light curves and resulting Corazon TCEs per searched TESS sector for both our samples are shown in Table \ref{tab:tglc_noise_transit}.

\begin{deluxetable*}{lcc|cc}
\tabletypesize{\scriptsize}
\tablewidth{0pt}
\tablecaption{The number of targets, TGLC light curves, and Corazon generated TCEs associated with our sample discussed in \S\ref{subsec:TESS}.
\label{tab:tglc_noise_transit}}
\tablehead{}
\startdata
\multicolumn{1}{l}{} &
\multicolumn{2}{c}{\textbf{No-KOI Control Sample}} &
\multicolumn{2}{c}{\textbf{Transit-like KOI Reference Sample}} \\
\cline{2-3} \cline{4-5}
\multicolumn{1}{l}{\textbf{Number of Targets}} &
\multicolumn{2}{c}{\num{177850}} &
\multicolumn{2}{c}{\num{7572}} \\
\hline
\multicolumn{1}{l}{\textbf{Sector}} &
\textbf{TGLC Light Curves} &
\textbf{Corazon TCEs} &
\textbf{TGLC Light Curves} &
\textbf{Corazon TCEs} \\
\hline
14    & \num{170634} & \num{219301} & \num{7306}  & \num{11012} \\
15    & \num{102457} & \num{119751} & \num{4497}  & \num{6422}  \\
16    & \num{166}    & \num{185}    & \num{12}    & \num{17}    \\
26    & \num{38124}  & \num{41260}  & \num{1334}  & \num{1740}  \\
\hline
Total & \num{311381} & \num{380497} & \num{13149} & \num{19191} \\
\enddata
\tablecomments{
The sum of light curves across our searched sectors is not the same as the number of unique targets since some targets were observed in multiple sectors and therefore generated multiple TGLC light curves. From this we also find an average of 1.2 and 1.5 Corazon-identified TCEs per TGLC light curve for the no-KOI and transit-like KOI targets. The higher rate of detections in the KOI target light curves is likely due to finding the primary and secondary of eclipsing binaries and counting those both as separate TCEs. 
}
\end{deluxetable*}

\subsection{Federation to KOIs} \label{subsec:federation}
We create a sample of transit-like reference signals by federating the TCEs on the KOI light curves against the KOI table ephemerides and dispositions. We consider a TCE to be federated if it satisfies three criteria. First, the ratio of the Corazon signal's period to the published KOI period must be within $\pm$0.1 of a multiple of 0.5. This ensures that Corazon detections of signals falling at half or double periods, such as eclipsing binaries, are not labeled as un-recovered while still requiring the measured period to align with the KOI value. Second, we require that more than 50\% of the Corazon-identified in-transit cadences are within the in-transit times given by propagating the KOI transit epoch, duration, and period to the start of the TGLC time series.

If a single Corazon-generated TCE passes both of these criteria for more than one associated KOI, we choose the KOI with the closest period and highest fraction of correct in-transit cadences. This only occurs for five out of the 19,191 TCEs. Finally, we require the matched KOI to have a ``Not Transit-Like False Positive Flag'' of 0 in the KOI table, i.e., we only federate to KOIs that are not likely to be from instrumental artifacts or low SNR detections. In total, we find that 1,976 out of the 19,191 TCEs from the transit-like KOI reference sample are federated to known transit signals, corresponding to 931 unique KOIs. 

\subsection{Setting Vetting Thresholds} \label{subsec:vetting}
\begin{figure*}
\centering
\includegraphics[width=0.9\textwidth]{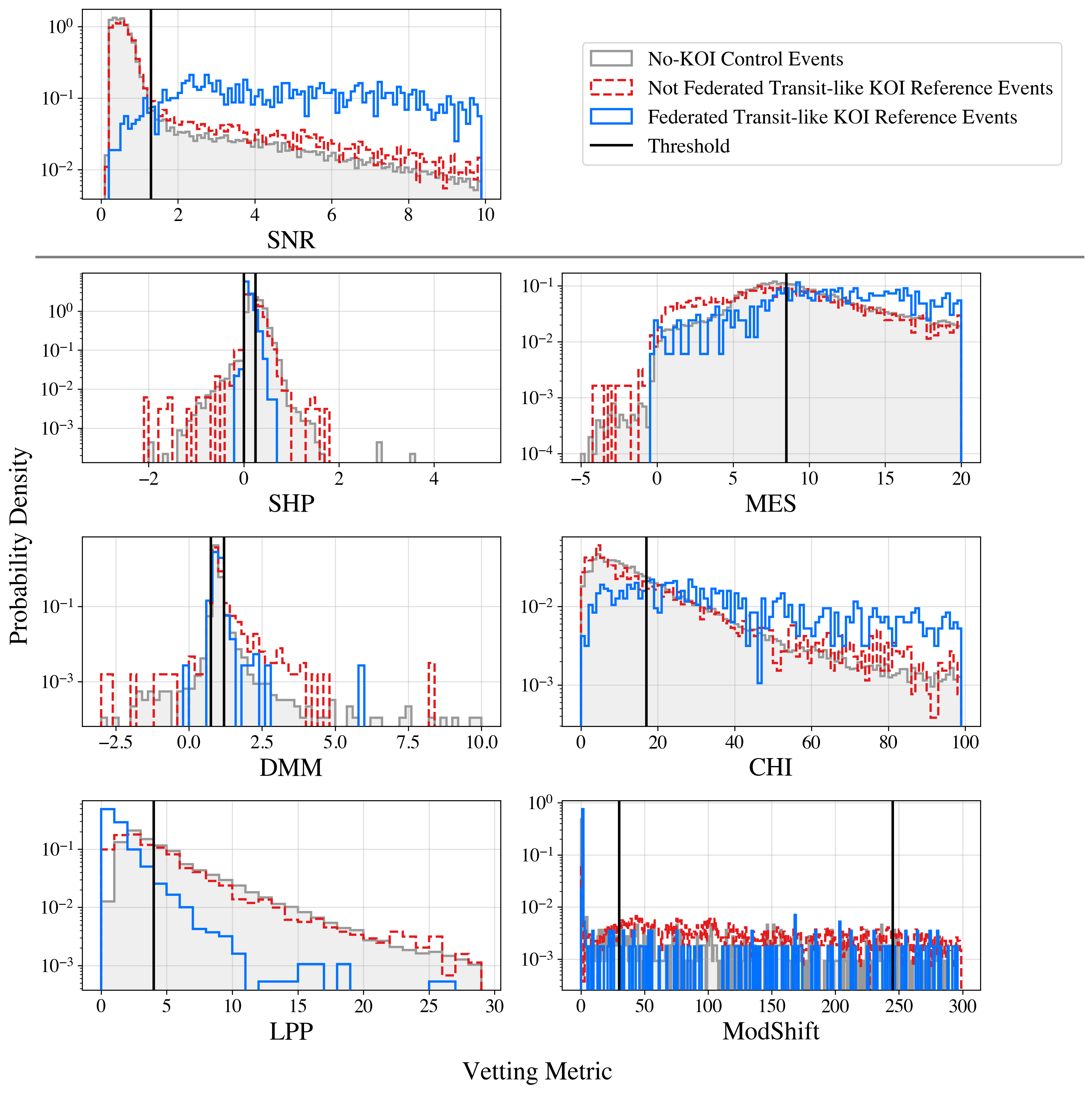}
\caption{Distributions of the chosen vetting metrics for the federated transit-like KOI reference events, not federated transit-like KOI reference events, and no-KOI control populations. The vetting metric distributions are shown across the TCEs which remained after applying an initial SNR $>$ 1.3 cut. The thresholds were chosen to include as many of the federated transit-like KOI TCEs as possible while removing as many of the not federated KOI and no-KOI control TCEs as possible. Note that the MES and ModShift metrics can return NaNs, which are not displayed, but account for a majority of the no-KOI control events flagged by these metrics.
\label{fig:vetting_distributions}}
\end{figure*}

We use \texttt{exoVetter} to evaluate each TCE in both the no-KOI control sample and the transit-like KOI reference samples as described in \S\ref{subsec:TESS}. Ideally, the federated subset of the transit-like KOI reference sample should pass all the vetters, and the not federated KOI events along with the no-KOI control events should fail at least one metric. In Figure~\ref{fig:vetting_distributions} we plot the distribution of vetting metrics for each sample. Because the number of not federated KOI events and no-KOI control events far exceeds the number of federated KOI events below an SNR of 1.3, we choose to only consider events above this SNR. Since the remaining sample is made up of known no-KOI control events and transit-like KOI events, we adjust the set of vetting thresholds to maximize the number of no-KOI control events removed while minimizing the number of federated transit-like KOI events removed. As these thresholds are selected and evaluated using the same sample, the reported values in \S\ref{subsec:effectiveness} are in-sample calibration results rather than independent performance estimates. Our thresholds are shown numerically in Table \ref{tab:vetting_cuts}. They are also overplotted as black lines in Figure~\ref{fig:vetting_distributions}. It is up to the user to specify the threshold of each vetting metric given their science goals. Users who are willing to accept more false alarms may choose to relax these thresholds while those who desire a more reliable final catalog, or are using \texttt{exoVetter} to distinguish between planets and eclipsing binaries, may need to run additional vetters (e.g., odd-even) provided in the \texttt{exoVetter} package.

\begin{deluxetable*}{lc|ccc}
\tabletypesize{\scriptsize}
\tablewidth{0pt}
\tablecaption{Summary of Vetting Cuts\label{tab:vetting_cuts}}

\tablehead{
\multicolumn{1}{l}{} &
\multicolumn{1}{c}{} &
\multicolumn{3}{c}{\textbf{TCEs Removed}} \\
\colhead{Vetting Metric} &
\colhead{Threshold} &
\colhead{Federated Transit-like KOI Events} &
\colhead{Not Federated Transit-like KOI Events} &
\colhead{No-KOI Control Events}
}
\startdata
SNR                       & $> 1.3$   & 82 & 12,610 & 307,611 \\
\tableline
SHP                       & $0 < \mathrm{SHP} < 0.25$   & 213 & 2,525 & 49,863 \\
MES                       & $> 8.5$                       & 205 & 2,619  & 44,962 \\
DMM                       & $0.75 < \mathrm{DMM} < 1.2$ & 113 & 1,621  & 28,096 \\
CHI                       & $> 17$ & 264 & 3,082  & 53,252 \\
LPP                       & $< 4$ & 138 & 2,066  & 37,275 \\
ModShift                  & $< 30$ or $> 245$ & 91 & 1,955  & 71,591 \\
\tableline
\textbf{Total}            &                             &
\textbf{549 (-27.78\%)}    &
\textbf{16,737 (-97.22\%)}&
\textbf{380,338 (-99.96\%)} \\
\enddata

\tablecomments{As in Figure \ref{fig:vetting_distributions} the number of TCEs removed by the vetters is taken after applying the initial SNR $>$ 1.3 cut shown in the first row. The total removed by applying the SNR and all vetting thresholds is not the same as the sum of the number removed from each individual metric because some TCEs are flagged by multiple vetters. The individual rows are present to give a sense of which metrics provide the largest cuts while the ``Total'' row shows the reduction in each population compared to the totals shown in Table \ref{tab:tglc_noise_transit}.}
\end{deluxetable*}

\subsection{Vetting Performance} \label{subsec:effectiveness}
Having selected thresholds for our vetting metrics, we apply them to our sample and analyze their performance. Similar to the completeness, effectiveness, and reliability analysis performed in Section 7.3 of \citet{Thompson2018}, we define our own completeness as the federated-KOI retention fraction, our own effectiveness as the no-KOI control-sample TCE rejection fraction, and our own reliability as the federated-KOI fraction among surviving TCEs to characterize our vetting:

\begin{itemize}
\item \textbf{Completeness (Federated-KOI retention fraction):} The fraction of federated transit-like KOI events remaining after vetting,
\begin{equation}
  \text{C} = \frac{\text{Remaining Federated Transit-like KOI Events}}{\text{Total Federated Transit-like KOI Events}} .
\end{equation}

\item \textbf{Effectiveness (No-KOI control-sample TCE rejection fraction):} The fraction of no-KOI control signals removed through vetting,
\begin{equation}
   \text{E} = \frac{\text{No-KOI Control TCEs removed}}{\text{Total No-KOI Control TCEs}} .
\end{equation}

\item \textbf{Reliability (Federated-KOI fraction among surviving TCEs):} The fraction of TCEs remaining after vetting that are reference transit-like KOIs,
\begin{equation}
    \text{R} = \frac{\text{Federated Reference Transit-like KOI TCEs}}{\text{Total vetted TCEs}} .
\end{equation}

\end{itemize}

\begin{figure*}
\centering
\includegraphics[width=\textwidth]{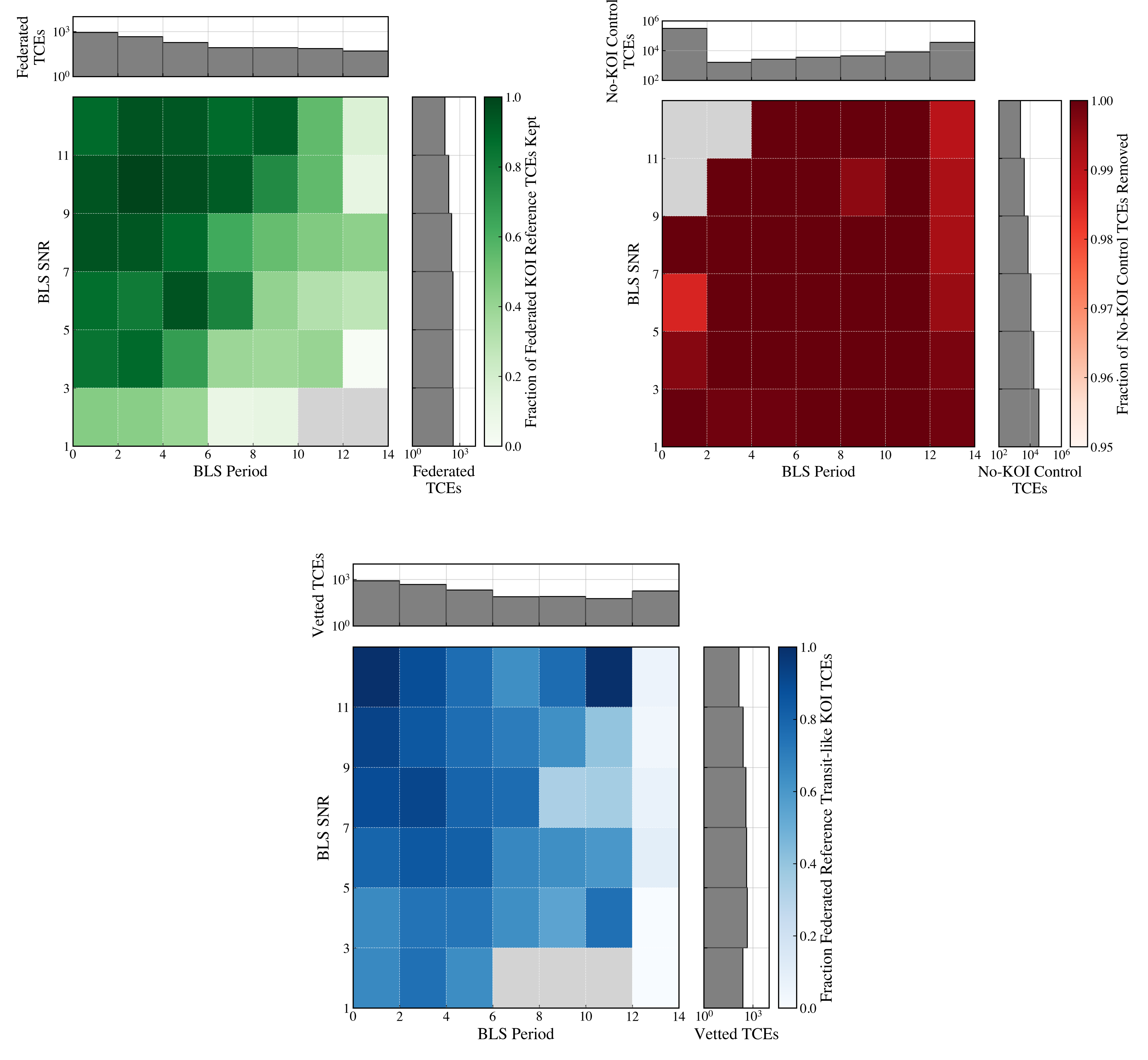}
\caption{Our vetting completeness (top left panel) and effectiveness (top right panel) of our initial static vetting cuts shown in Figure \ref{fig:vetting_distributions} and Table \ref{tab:vetting_cuts} when applied to the federated and no-KOI control TCEs, respectively. The reliability of the resulting federated events is shown in the bottom panel. For visual clarity we only plot SNRs from 1 to 13 as we automatically reject TCEs with SNR $\leq 1.3$ and a majority of both the no-KOI control and federated transit-like KOI TCEs have SNRs $< 13$. We also display regions with $<$ 5 total federated transit-like KOI events prior to vetting as gray in the completeness plot, $<$ 5 total no-KOI control events prior to vetting as gray in the effectiveness plot, and $<$ 5 total events post vetting as gray in the reliability plot.
\label{fig:completeness_effectiveness}}
\end{figure*}

In Figure \ref{fig:completeness_effectiveness} we show the federated-KOI retention fraction, no-KOI control-sample TCE rejection fraction, and federated-KOI fraction among surviving TCEs of our sample binned across period and SNR space. 

Across all periods and SNRs $>1.3$ we achieve $C \approx 75\%$, $E \approx 99.8\%$, and $R \approx 69\%$. This is improved to $C \approx 89\%$, $E \approx 99.9\%$, and $R \approx 81\%$ when we limit to periods $\leq12$ days and SNRs $>7$. For a real planet search, further improvements can be made to the vetting performance by including more vetting metrics and/or providing different thresholds in different regions. One of the benefits of \texttt{exoVetter} is that the light curves tested, the metrics used, and the thresholds chosen in a pipeline can be tailored for each individual science case. 

We summarize vetting performance in different regions of SNRs and periods in Table \ref{tab:summary}. Included in Table \ref{tab:summary} alongside our completeness, effectiveness, and reliability is the number of remaining no-KOI control TCEs in each region per total input no-KOI control light curve. We also show the performance of the vetters on ``bright'' targets $\le 13.5$ \tmag\ and ``dim'' targets $> 13.5$ \tmag. Having extracted light curves on dim targets is one of the primary advantages of TGLC, as QLP \citep{Huang2020} and SPOC FFI \citep{SPOC} products only go down to 13.5 \tmag. 

\begin{deluxetable*}{lcccc}
\tabletypesize{\scriptsize}
\tablewidth{0pt}
\tablecaption{Summary of Vetting Performance\label{tab:summary}}
\tablehead{
\colhead{} &
\colhead{Vetting Completeness} &
\colhead{Effectiveness} &
\colhead{Reliability} &
\colhead{False Alarms per Light Curve}
}
\startdata
BLS SNR $> 1.3$ & $0.75\ (1,427/1,894)$ & $9.978\times 10^{-1}\ (72,727/72,886)$ & $0.69\ (1,427/2,064)$ & $5.1\times 10^{-4}\ (159/311,381)$ \\
\tableline
High SNR, Short Period & $0.89\ (665/748)$ & $9.998\times 10^{-1}\ (8,989/8,991)$ & $0.81\ (665/824)$ & $2.1\times 10^{-4}\ (65/311,381)$ \\
High SNR, Short Period, Bright & $0.88\ (365/414)$ & $9.993\times 10^{-1}\ (1,357/1,358)$ & $0.82\ (365/447)$ & $3.2\times 10^{-5}\ (10/311,381)$ \\
High SNR, Short Period, Dim & $0.90\ (300/334)$ & $9.999\times 10^{-1}\ (7,632/7,633)$ & $0.80\ (300/377)$ & $1.8\times 10^{-4}\ (55/311,381)$ \\
\enddata
\tablecomments{We define high SNR as $>7$, short period as $\leq$12d, bright as ($\leq$13.5 \tmag), and dim as ($>$13.5 \tmag). In this context ``completeness'' refers to the fraction of federated transit-like KOI events
remaining after vetting, ``effectiveness'' refers to the fraction of no-KOI control signals removed through vetting, ``reliability'' refers to the fraction of TCEs remaining after vetting that are reference transit-like KOIs, and ``False Alarms per Light Curve'' refers to the number of remaining control No-KOI TCEs per input control No-KOI light curve.}
\end{deluxetable*}

Our analysis of the search on dim and bright targets shows that the TGLC light curves can be used for planet searches on dim targets. We find that although a majority of the short-period, high-SNR TCEs from the no-KOI control sample are generated on dim targets, we are able to achieve almost identical completeness and effectiveness (see Table \ref{tab:summary}), along with a comparable reliability for events of high SNR and periods less than 12 days. Our ability to distinguish noise from real transits on dim targets with TGLC data empirically demonstrates how the light curves generated by \citet{Han2023AJ} will allow the community to extend TESS FFI transit searches to dimmer stars without a large increase in false positives. We note that because of the success of the TGLC algorithm for dimmer targets, the TESS Quick Look Pipeline \citep{Kunimoto2022b, QLP_04} began using TGLC photometry starting in sector 94.

\subsection{Comparing Centroid Vetting} \label{subsec:centroid}

We explore the ability of the \texttt{exoVetter} centroid offset detection in TESS data by comparing to known centroid offsets from Kepler. We federate 807 KOIs with sufficiently short periods to produce at least 2 transits in the TESS data where we can make a good centroid measurement from \texttt{exoVetter}. Kepler identifies a significant centroid offset on 281 of these KOIs \citep{Thompson2018}. We run the centroid vetter on each of these KOIs individually and consider a significance value $<$ 0.003 as indicative of a real offset. We find that the \texttt{exoVetter} run on TESS data correctly identifies 162 of the 281 KOIs as also having a significant offset ($\approx58\%$). \texttt{exoVetter} performs slightly better on targets brighter than 13.5 \tmag, correctly identifying 47 out of the 70 bright KOIs $\leq$ 13.5 \tmag\ while identifying 115 out of the 211 dim KOIs $>$ 13.5 \tmag, which is consistent with our findings in \S\ref{subsec:effectiveness}.
Because of the larger pixel size and the smaller collecting area of the mirror on TESS (21$''/$pixel and 10\,cm vs.~$3.\!\!''98/$ pixel and 0.95\,m for Kepler), a centroid offset found in Kepler would need to be quite separated from the star for it to be detectable in TESS. It is therefore not surprising that we do not recover all centroid offsets found by Kepler. 

\begin{figure*}
\centering
\includegraphics[width=0.6\textwidth]{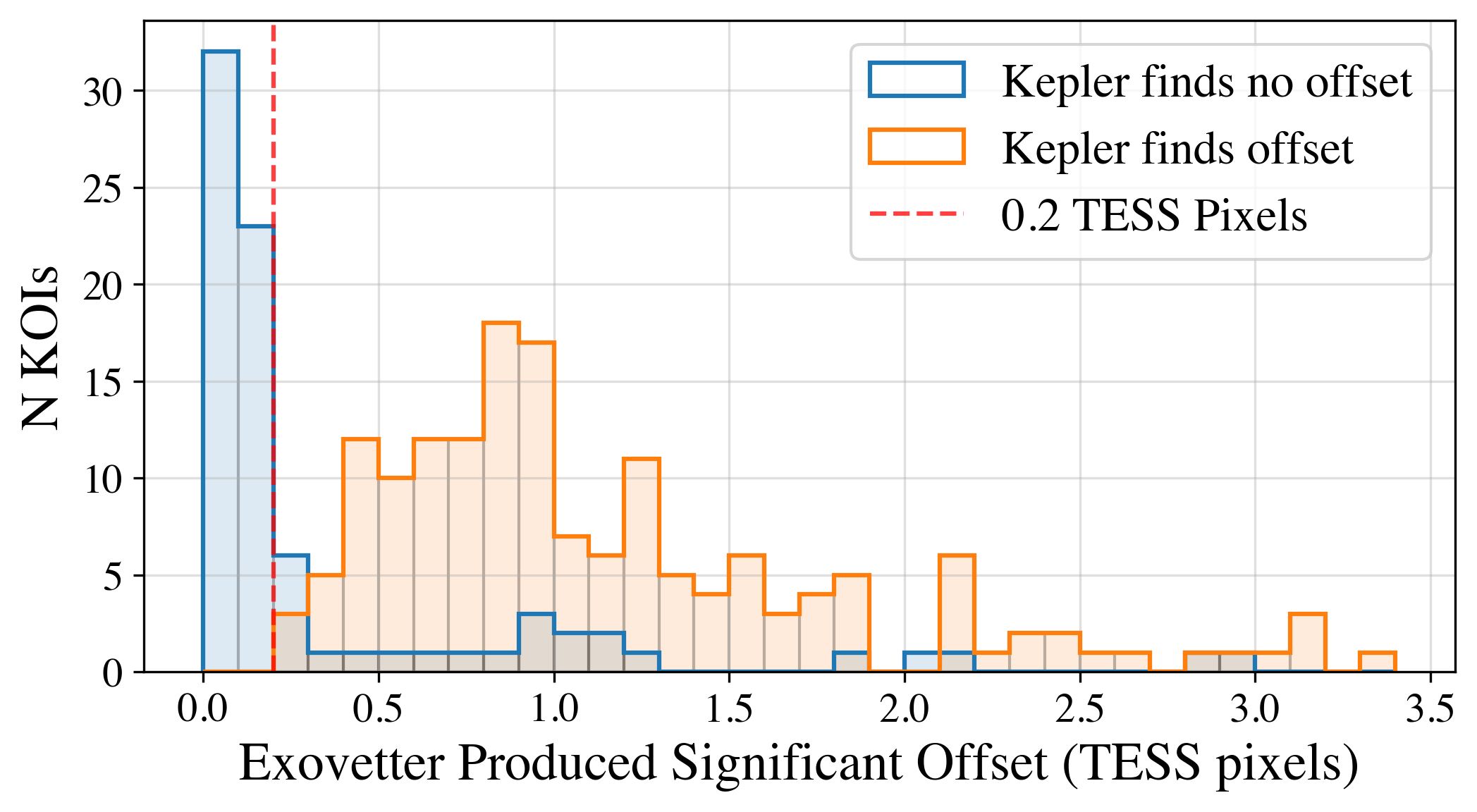}
\caption{The distribution of offsets on KOIs that \texttt{exoVetter} incorrectly identifies as having a significant centroid offset from TESS TPFs (blue), and correctly identifies with a significant centroid offset (orange). If a KOI was observed multiple times (i.e., in different sectors) we take the median offset. A majority of these spurious centroid offsets are $<$ 0.2 TESS pixels, smaller than TESS can reasonably be expected to identify. 
\label{fig:offset_dist}}
\end{figure*}

We also assess whether \texttt{exoVetter} finds centroid offsets on signals where we expect no known offset. Kepler, with its smaller pixels, should more precisely identify the location of the transit on the sky. Of the 807 KOIs we tested, there are 27 listed in the Kepler Object of Interest table that are both planet candidates under the ``Disposition Using Kepler Data'' column and on target under the ``Centroid Offset False Positive Flag'' column. The \texttt{exoVetter} centroid algorithm displays a significant offset on one of the 27, planet candidate KOI-1154.01. 
In this instance the significance value is $<$ 0.003 but the measured offset itself is $<$ 0.1 TESS pixels. \texttt{exoVetter} can occasionally identify a very small, yet significant offset that is spurious. In our analysis we find that a majority of the incorrectly identified significant ``offsets'' found on KOIs (both false positives and planet candidates) are $<$ 0.2 TESS pixels. We show the distribution of these incorrect offsets on KOIs in Figure~\ref{fig:offset_dist}. Only 26 of the 188 ($\approx14\%$) \texttt{exoVetter}-identified significant offsets that are $>$ 0.2 TESS pixels are not known offsets from Kepler. Furthermore, the generated difference images and significance ellipsoid plots (as shown in Figure \ref{fig:centroid_offset_example}) often help determine whether the measured offset is reliable. We therefore encourage users to view the difference and significance plots to diagnose their offsets. Users may also find that adjusting the size of their adopted TESS cutout \citep{astrocut2019} to something smaller than the $5\times 5$~pixel default will prevent the out-of-transit centroid measurements from being pulled by residuals from nearby bright stars. 

\subsection{Cloud Implementation of the Transit Search} \label{subsec:cloud}

MAST hosts the TGLC data in the AWS Repository of Open Data\footnote{\url{https://registry.opendata.aws/collab/stsci/}}. These data are freely open to use with no cost to the user. We decided to implement our search using AWS compute proximate to the data to make use of their extensive array of compute options without first having to download or transfer data. While our search is modest in size ($\approx 200$ GB of data) and could have been performed on a typical laptop with about a week of run time, this implementation will allow us to expand the workflow to larger samples in the future without purchasing new computers. Using cloud resources, the final run was $\sim$20 minutes of wall clock time and $\sim$180 CPU-hours, and required only minimal knowledge about AWS services. This section summarizes our methods for full reproducibility of our software stack while also providing an example of how the cloud can be used to explore new science projects that require computing resources beyond what is available at researchers' own institutions.

A transit search is by nature an extremely parallel task since the searching and vetting of each light curve is entirely independent of the others. AWS provides many ways to run simple parallel tasks. Because of its user-friendly interface for Python users, we decided to use a third-party software called \texttt{Coiled}\footnote{\url{https://coiled.io}} (v 1.91.0) to create a Dask cluster of hundreds of workers in the AWS east-1a region, which is physically adjacent to the data. We saved compute time by turning off the plot generation parts of the exoVetter code during our run. Our final run of \nlcs\ light curves was executed on 480 \texttt{m6i.large} workers. It was accomplished in under 20 minutes and cost less than 7 US dollars. We were able to speed up the search primarily by parallelizing the task on a large number of AWS workers and also to a lesser degree by not having to first transfer data to our local computer. 

The cloud run required a few small changes to our Corazon pipeline to allow it to write the results to an AWS S3 bucket instead of to our local computer. To ensure the pipeline was entirely operating in the cloud, we created our own AWS S3 bucket for storing the outputs of our run and used the s3fs (v2024.6.1) package to mount our AWS S3 bucket like a local Unix file system so we could write to the bucket as we would normally write a file to our computer. When the run was completed, we continued our analysis in the cloud by making use of the TIKE science platform \citep{TIKE} provided by MAST to access our personal S3 buckets and do additional analysis using \texttt{lightkurve}, \texttt{astroquery} and \texttt{exoVetter} to spot check our results and do further analysis on the full table of results.

\section{Conclusions} \label{sec:conclusion}
In this paper we have introduced a transiting exoplanet vetting Python package called \texttt{exoVetter}\footnote{\url{https://github.com/spacetelescope/exovetter}}. The package allows users to run vetting metrics developed by different teams with a common programming interface and workflow and is designed to allow additions, as needed, to adapt vetting to new research projects. \texttt{exoVetter} is open source and welcomes well-tested additions to the package. Both flux and pixel-level vetting are supported to look for common sources of false alarms from instrumental noise, binary stars, variable stars, and blended background binaries.

To demonstrate a basic use case of \texttt{exoVetter} to classify a set of signals found in the TESS data, we perform an example transit search and vetting for over 300,000 TGLC light curves of targets observed by Kepler. With ground truth regarding what is a transit-like signal from Kepler, we are able to evaluate the performance of the \texttt{exoVetter} package on events returned by a BLS on these TESS light curves. With minimal tuning of thresholds on the metrics, for SNR $>7$ and periods $\leq$12 days, our adopted thresholds retain $89\%$ of federated KOI events, reject $99.9\%$ of TCEs in the no-KOI control sample, and produce a surviving sample in which $\sim81\%$ federate to known transit-like KOIs. Our sample of recovered Kepler transits in the TESS data also allows us to evaluate the performance of \texttt{exoVetter}'s centroid vetter. We find that \texttt{exoVetter} correctly identifies a significant offset on $\approx58\%$ of the KOIs with known offsets we tested against. We also find that the \texttt{exoVetter} centroid vetter can commonly return erroneous, significant offsets under 0.2 TESS pixels and should not be trusted to identify offsets of this size. However, of the \texttt{exoVetter}-identified offsets that are larger than 0.2 TESS pixels, only $\approx14\%$ of them are not associated with a known KOI offset. In most of these cases the diagnostic plots reveal low quality difference images or crowding that caused the false measurement.

Finally, since we have ground truth from Kepler, we compare the results of a transit search on targets brighter than 13.5\tmag\ and dimmer than 13.5\tmag. Despite the larger number of events found around the dimmer targets at low SNR, the \texttt{exoVetter}-based classifier can easily vet the extra signals resulting in similar federated-KOI fractions among surviving TCEs for short-period planets. This demonstrates how the TGLC light curves open up the possibility for an improved transit search around dimmer targets. 

\begin{acknowledgments}
This paper includes data collected with the TESS mission, obtained from the MAST data archive at the Space Telescope Science Institute (STScI). Funding for U.S. Institutions for the TESS mission is provided by the NASA Explorer Program. STScI is operated by the Association of Universities for Research in Astronomy, Inc., under NASA contract NAS 5–26555. The authors thank the Astrophysics Data Analysis Program for funding a portion of this effort as part of award number 80NSSC21K0631. MK acknowledges the support of the Natural Sciences and Engineering Research Council of Canada (NSERC), RGPIN-2024-06452. Cette recherche a été financée par le Conseil de recherches en sciences naturelles et en génie du Canada (CRSNG), RGPIN-2024-06452. The authors also thank the referee for their careful review, which provided useful suggestions that improved this work. 
\end{acknowledgments}

\facilities{TESS, Kepler}
\software{astropy \citep{astropy2013,astropy2018,astropy2022}, lightkurve \citep{lightkurve}, astrocut \citep{astrocut2019}, scipy \citep{scipy}, matplotlib \citep{Hunter2007}, numpy \citep{Harris2020}, exovetter \citep{exovetter}}

\bibliography{bibliography}{}
\bibliographystyle{aasjournalv7}

\end{document}